\documentclass[aps,prl,preprintnumbers,amsmath,amssymb,latexsym,nofootinbib,array,enumerate,letter,twocolumn]{revtex4-1}
\pdfoutput=1
\usepackage{graphicx}
\usepackage{epstopdf}
\usepackage{dcolumn}
\usepackage{bm}
\usepackage{hyperref}
\usepackage{color}
\usepackage{amsmath}
\usepackage{scalerel}
\usepackage{url}
\usepackage{ulem}
\usepackage{lipsum}
\usepackage{mleftright}
\mleftright

\usepackage{lineno}

\def\bino{\tilde{B}}
\def\tbino{\tilde{B}'}

\begin{document}

\title{Natural Twin Neutralino Dark Matter}

\author{Marcin Badziak, Giovanni Grilli di Cortona}
\affiliation{Institute of Theoretical Physics, Faculty of Physics, University of Warsaw, ul.~Pasteura 5, PL--02--093 Warsaw, Poland}
\author{Keisuke Harigaya}
\affiliation{School of Natural Sciences, Institute for Advanced Study, Princeton, New Jersey 08540, USA}

\begin{abstract}
Supersymmetric Twin Higgs models have a discrete symmetry for which each Standard Model particle and its supersymmetric partner have a corresponding state that transforms under a mirror Standard Model gauge group. This framework is able to accommodate the non-discovery of new particles at the LHC with the naturalness of the electroweak scale. We point out that supersymmetric Twin Higgs models also provide a natural dark matter candidate. We investigate the possibility that a twin bino-like state is the Lightest Supersymmetric Particle and find that its freeze-out abundance can  explain the observed dark matter abundance without fine-tuning the mass spectrum of the theory. Most of the viable parameter space can be probed by future dark matter direct detection experiments and the LHC searches for staus and higgsinos which may involve displaced vertices.
\end{abstract}

\maketitle

{\it Introduction.}---%
For decades the primary motivations for supersymmetry (SUSY), in particular the Minimal Supersymmetric Standard Model (MSSM), have been a solution to the hierarchy problem~\cite{Maiani:1979cx,Veltman:1980mj,Witten:1981nf,Kaul:1981wp} and the Lightest Supersymmetric Particle (LSP) as a natural dark matter (DM) candidate~\cite{Witten:1981nf,Pagels:1981ke,Goldberg:1983nd}. However, recent experimental results show that the MSSM no longer provides a natural solution to the hierarchy problem, and the LSP can be DM only in fine-tuned corners of the parameter space.

The LHC has found the Higgs boson with a mass about 125 GeV~\cite{Aad:2012tfa,Chatrchyan:2012xdj}, which requires heavy stops~\cite{Okada:1990gg,Okada:1990vk,Ellis:1990nz,Haber:1990aw}. As a result, the electroweak scale is obtained only by at least a permille level of fine-tuning of parameters. Lighter stops can accommodate the Higgs mass in extension of the MSSM with a singlet field as in the Next-to-MSSM (NMSSM)~\cite{Fayet:1974pd,Nilles:1982dy,Frere:1983ag,Ellwanger:2009dp,Hall:2011aa,Ellwanger:2011aa}, electroweak charged fields with large Yukawa couplings to the Higgs~\cite{Moroi:1991mg,Moroi:1992zk,Martin:2009bg,Endo:2011mc,Moroi:2011aa}, or a new gauge interaction under which the Higgs is charged~\cite{Langacker:1999hs,Batra:2003nj,Morrissey:2005uz,Cheung:2012zq}. Although these models can easily accommodate the Higgs mass, they are still fine-tuned at least at a percent level due to lower bounds on stop and gluino masses from direct LHC searches~\cite{Aaboud:2017ayj,ATLAS-CONF-2019-040,CMS:2019twi}. The necessity of the fine-tuning is called the little hierarchy problem.

Among the scenarios of the LSP DM, the neutralino LSP with its abundance determined by the freeze-out~\cite{Lee:1977ua} is particularly interesting because of predictability and possible signals in (in)direct detection experiments. 
In the MSSM there are four neutralinos -- bino, wino and two higgsinos -- and each of them may potentially play the role of the DM. 
However, nowadays neutralino DM in the MSSM is either excluded or requires fine-tuning. 
Pure wino or higgsino LSP have large annihilation cross-section and their relic abundance is obtained only for large DM mass of about 3 or 1 TeV, respectively~\cite{Hisano:2006nn}. They are consistent with experimental constraints, but requiring the SUSY mass scale above TeV is inconsistent with the natural electroweak scale.
The bino-higgsino mixed LSP allows for $\mathcal{O}(100)$ GeV soft masses~\cite{ArkaniHamed:2006mb}, but is excluded by recent DM direct-detection (DD) experiments~\cite{Badziak:2017the}. A pure bino is experimentally viable, but the correct abundance $\Omega h^2~\approx~0.12$~\cite{Aghanim:2018eyx} is obtained only for a fine-tuned mass spectrum enabling coannihilation or resonant annihilation~\cite{Griest:1990kh}.

The little hierarchy problem can be solved by the Twin Higgs (TH) mechanism~\cite{Chacko:2005pe}.
The $\mathbb{Z}_2$ symmetry introduced in the TH mechanism relaxes the fine-tuning and predicts a mirror copy of the Standard Model (SM) particles which we denote by superscripts $'$.
Recently a new class of SUSY
TH models was proposed~\cite{Badziak:2017syq,Badziak:2017wxn,Badziak:2017kjk} which naturally predicts the observed Higgs mass and allows for tuning of the electroweak scale at the level of $\mathcal{O}(10)\%$ even for stops and gluino masses above 2~TeV.
Comparable or even larger amount of tuning was required in the MSSM already after LEP~\cite{Chankowski:1997zh,Barbieri:1998uv} while the MSSM tuning is currently at a permille level. This is a strong motivation to look more closely at phenomenological aspects of SUSY TH models.

In this Letter, we point out that SUSY TH models also provide a natural DM candidate. The lightest twin neutralino tends to be lighter than the corresponding MSSM neutralino and may be the LSP with correct relic abundance.
Because of the TH mechanism the higgsino mass may be much above the electroweak scale. In that case, the twin bino-like state $\tbino$ is a natural candidate for the LSP. 
The key observation is that $\tbino$ annihilates into twin fermions without chirality suppression if the twin fermions are heavier than the SM ones.  Heavy twin fermions are well-motivated since they allow to avoid excessive dark radiation~\cite{Barbieri:2016zxn,Barbieri:2017opf,Harigaya:2019shz}, and make TH models compatible with cosmological constraints~\cite{Aghanim:2018eyx}. We demonstrate salient features of twin neutralino DM being a mixture of the twin bino and higgsino and compare to the bino-higgsino LSP in the MSSM which has been exhaustively studied in the literature~\cite{ArkaniHamed:2006mb,Baer:2006te,Farina:2011bh,Cheung:2012qy}. A detailed study of the LSP being a mixture of all twin neutralinos will be presented elsewhere.

We list other DM candidates in TH models.
Twin neutrons and twin neutral atoms, and if the twin electromagnetic symmetry is broken, twin electrons, may be DM~\cite{Farina:2015uea,Barbieri:2016zxn,Prilepina:2016rlq,Barbieri:2017opf,Koren:2019iuv}. In fraternal TH~\cite{Craig:2015pha}, twin taus, mesons and bottom baryons may be also DM~\cite{Garcia:2015loa,Garcia:2015toa,Craig:2015xla,Farina:2015uea,Freytsis:2016dgf,Prilepina:2016rlq,Hochberg:2018vdo,Terning:2019hgj}. They enjoy phenomenology such as self-interacting DM and direct detection.
We find that the twin bino-like LSP also has rich phenomenological consequences including signals at nuclear recoil experiments and the LHC. We work in the framework of mirror TH models but twin neutralino LSP can be a good DM candidate also in other variants of TH models such as fraternal TH.

{\it Thermal abundance of Twin Bino LSP}---%
We focus on a twin bino-dominated LSP. We assume for simplicity that
the twin wino is decoupled. This assumption does not affect our results unless the relative bino-wino mass splitting is less than several tens of percent. On the other hand, even with the TH mechanism, naturalness requires that higgsinos are relatively light. The LSP has some twin higgsino component with non-negligible impact on observables.
We also assume that the twin and MSSM neutralinos mix with each other by a small amount, so that the MSSM bino $\bino$ can decay into $\tbino$ before the freeze-out of $\tbino$ occurs, which is the case in typical models~\cite{Falkowski:2006qq,Chang:2006ra,Badziak:2017syq,Badziak:2017wxn,Badziak:2017kjk,Craig:2013fga,Katz:2016wtw}. 

\begin{figure}[t]
 \includegraphics[width=0.48\textwidth]{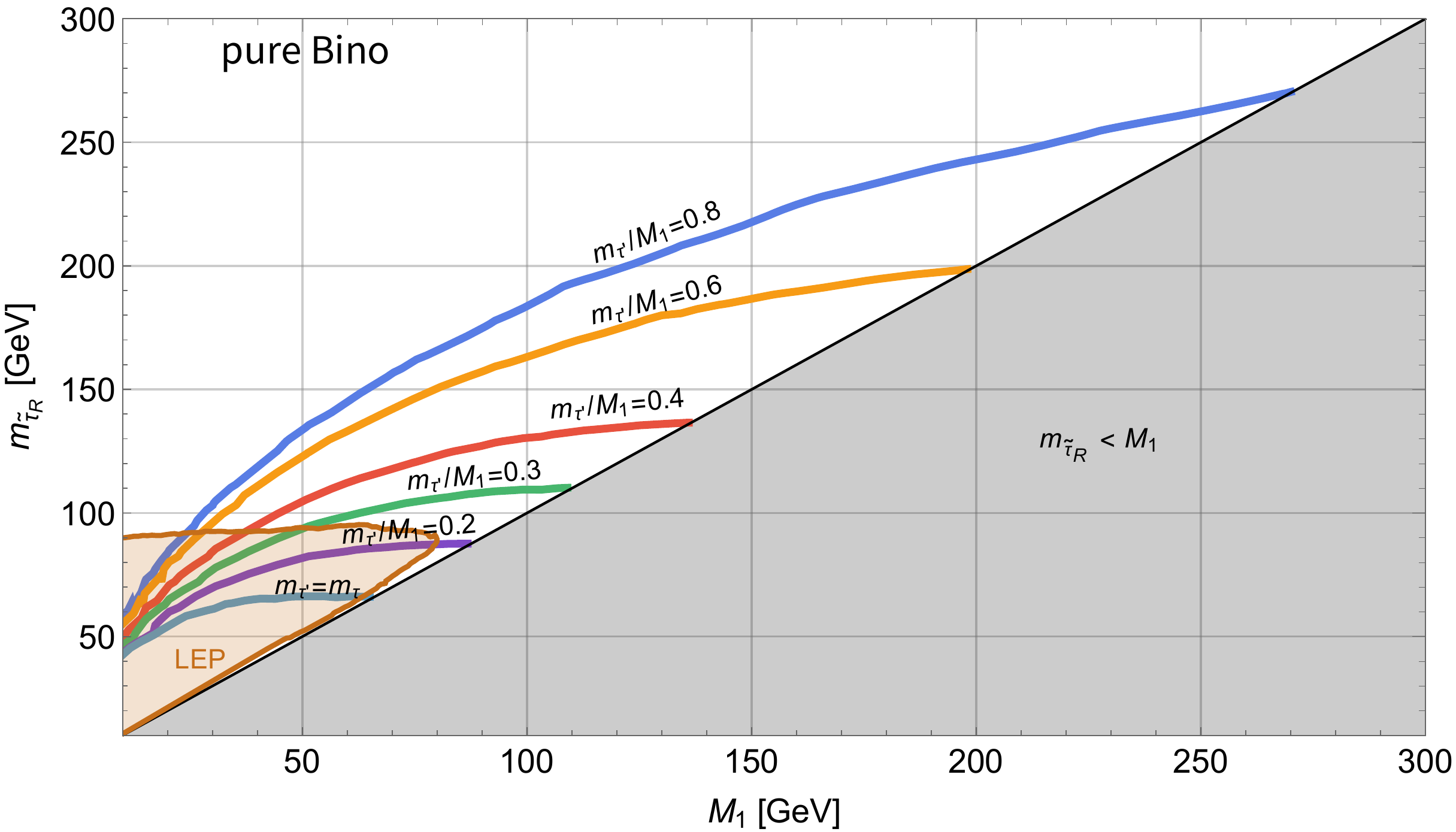}
 \caption{Contours of $\Omega h^2=0.12$ for pure twin bino DM in the $(M_1,m_{\tilde \tau_R})$-plane for several values of the ratio $m_{\tau'}/M_1$. The grey line with $m_{\tau'}=m_{\tau}$ corresponds to the pure bino DM in the MSSM. The effect of coannihilations is not included.   }
\label{fig:Omega}
\end{figure} 

Pure (twin) binos annihilate into (twin) fermions via t-channel exchange of (twin) sfermions $\tilde{f}$ with the cross-section scaling as $M_{1}^2/m_{\tilde f}^4$, where $M_1$ is the bino mass. Due to collider constraints on sfermion masses from LEP and the LHC this cross-section is suppressed, leading to $\bino$ LSP overabundance. Among sfermions,  the right-handed stau mass is least constrained due to its small pair production cross-section and $\tau$s in the final state. LEP sets a lower bound on the right-handed stau mass of about 90~GeV~\cite{Heister:2001nk,Abdallah:2003xe,Achard:2003ge,Abbiendi:2003ji}. Dedicated searches by ATLAS~\cite{ATLAS:2019ucg} with $139$~fb$^{-1}$ and CMS~\cite{CMS:2019eln} with $77$~fb$^{-1}$ of data, do not improve this bound. However, even for a stau mass of 100~GeV the relic abundance of binos is too large unless co-annihilations with staus are efficient, which requires fine-tuning of the mass splitting between the bino and the stau~\cite{Griest:1990kh,Ellis:1998kh}.

The annihilation cross-section of $\tbino$ is enhanced if there is a $\mathbb{Z}_2$ breaking in the Yukawa couplings such that the twin fermions are heavier than the corresponding SM ones. This avoids the chirality suppression of the s-wave amplitude which is present in the MSSM due to the fermion mass $m_f~\ll~M_{1}$.

The $\mathbb{Z}_2$ breaking in the Yukawa couplings is mandatory in order to avoid excessive dark radiation~\cite{Barbieri:2016zxn} in generic TH models unless a non-standard cosmological evolution is assumed~\cite{Craig:2016lyx,Chacko:2016hvu,Csaki:2017spo}.
The $\mathbb{Z}_2$ breaking does not introduce fine-tuning of the electroweak scale as long as there is no $\mathbb{Z}_2$ breaking in the top Yukawa coupling and the rest of the twin fermions are much lighter than the twin top quark. On the contrary, the $\mathbb{Z}_2$ breaking in the Yukawa couplings can radiatively generate
the difference between the electroweak scale $v$ and the twin electroweak scale $v'$, such that $v'>v$. 
A spontaneous $\mathbb{Z}_2$ breaking can happen in a relatively simple way~\cite{Barbieri:2017opf} by adapting the Froggatt-Nielsen mechanism~\cite{Froggatt:1978nt}.  

We consider the minimal case of the $\mathbb{Z}_2$ breaking in the Yukawa couplings and assume that the masses of MSSM and twin sparticles are nearly degenerate with each other, up to corrections given by the Yukawa couplings.

In our numerical computations we assume a simplified model in which the right-handed stau is the only light sfermion. Such a simplified model approximates well the relevant part of the spectrum in many models of SUSY breaking. Indeed the effects of renormalization-group running generically make the right-handed stau the lightest sfermion unless it is heavier than other sfermions at the mediation scale of SUSY breaking. Even in the presence of other light sfermions the results would be unaffected unless the corresponding twin fermion masses are similar to the twin tau mass. In this case the annihilation cross-section of $\tbino$ would be further enhanced making it even easier to obtain the correct relic abundance.
 
In figure~\ref{fig:Omega} we present contours of $\Omega h^2=0.12$
for a pure twin bino LSP for several values of the twin tau to bino mass ratio. An analogous contour for twin tau degenerate with the SM tau is also presented for comparison with the MSSM case.  In contrast to the MSSM case, the correct relic abundance can be obtained for a large range of bino and stau masses in agreement with the LEP constraints and without invoking co-annihilation. This happens if the twin tau mass is at least one third of the bino mass.

{\it Constraints and prospects for direct detection}---%
A pure twin bino has a small scattering cross-section with nuclei, far below the irreducible neutrino background for DD.%
\footnote{A pure MSSM bino also has small DD cross-section but it may be above the neutrino background when the mass splitting with the Next-to-LSP is below ${\mathcal O}(10)$~GeV~\cite{Berlin:2015njh}.}
However, naturalness in TH models requires that the $\mu$ parameter, which sets the masses of the higgsinos and their twins, cannot be arbitrarily large.
Some mixing between the twin higgsino and the twin bino is unavoidable, leading to LSP scattering off nuclei via the tree-level Higgs portal.
The effective Higgs coupling to the twin bino-like LSP $\tilde{B}'$ reads, modifying the results for the MSSM in~\cite{Badziak:2015qca},
\begin{equation}
\label{eq:chLSP}
{\cal L} = \frac{1}{2} c_{h}\, h \tilde{B}'\tilde{B}' + {\rm h.c.},~
c_{h}
\approx
\frac{v}{v'}
\frac{g_1^2 v'}{2\sqrt{2}\mu}
\,\left(s_{2\beta}
+\frac{M_1}{\mu}\right) \,,
\end{equation}
where $s_{2\beta}$ stands for $\sin2\beta$, with $t_\beta\equiv\tan\beta$ being the ratio of the vacuum expectation value (vev) of the up-type Higgs to that of the down-type Higgs, and $g_1$ is the hypercharge gauge coupling.
In the leading approximation the coupling does not depend on $v'$. Indeed, the suppression of the coupling by the mixing between the twin and SM-like Higgses (which is the portal to the twin sector), corresponding to the factor of $v/v'$ on the right-hand side,  is compensated by the enhanced twin bino-higgsino mixing which is proportional to $v'$.  
Nevertheless, the spin-independent (SI) DD cross-section (which is mediated by the SM-like Higgs) is expected to be much smaller than in the MSSM because the TH mechanism allows for much larger values of $\mu$ compatible with naturalness. In some UV completions of the TH model $\mu$ as large as 1 TeV leads to mild tuning at the level of only 10\%~\cite{Badziak:2017syq,Badziak:2017wxn,Badziak:2017kjk}.  

\begin{figure}[t]
\begin{center}
    \includegraphics[width=0.48\textwidth]{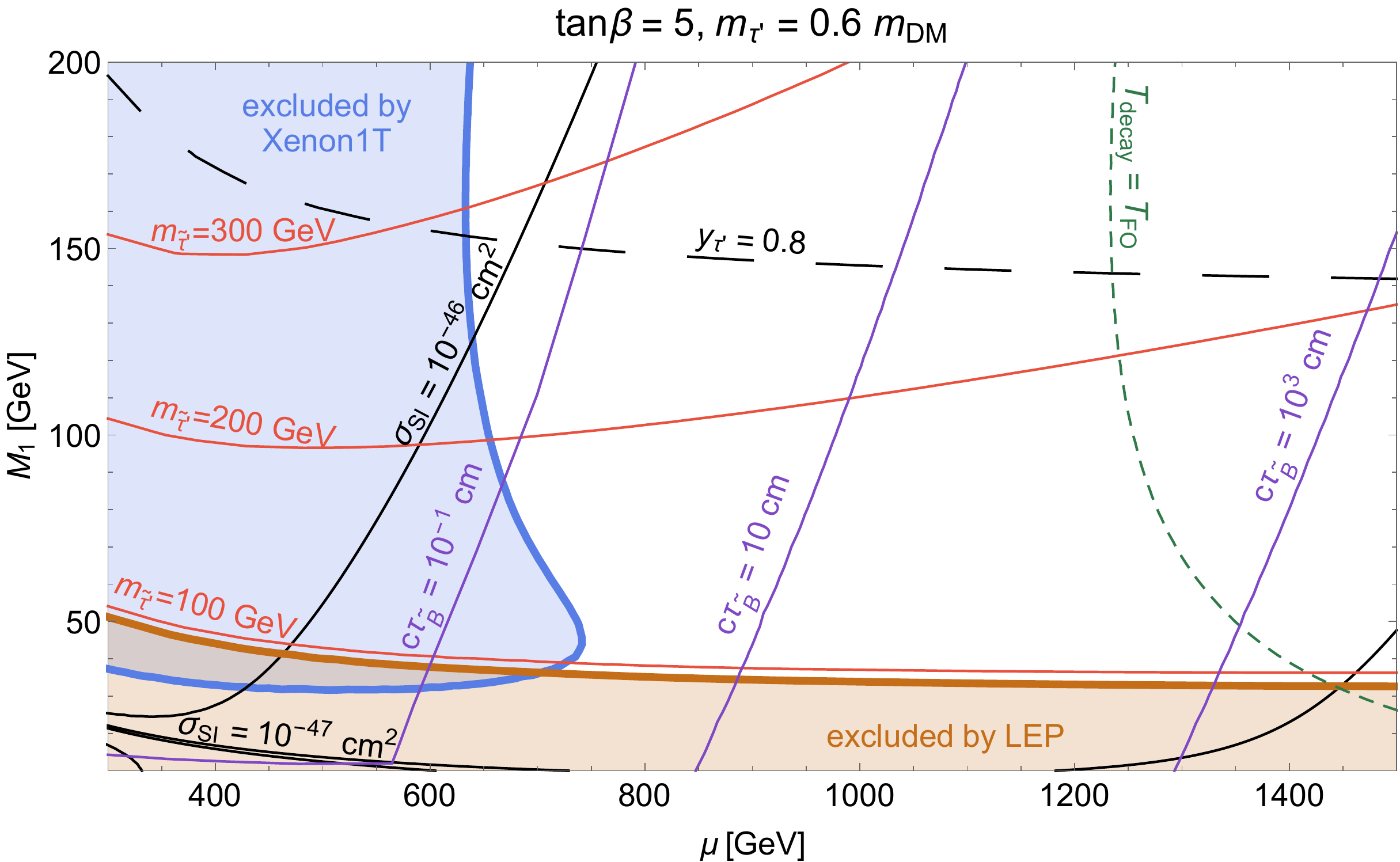}
  \end{center} 
\caption{SI DD cross-section (black) in the $(\mu,\,M_1)$-plane for $\tan\beta=5$, $v'/v=3$ and $m_{\tau'}/m_{\rm DM}=0.6$. The blue-shaded region is excluded by Xenon1T~\cite{Aprile:2018dbl}. 
The value of $m_{\tilde{\tau}'}$ reproducing the correct relic density of  DM is shown by the red contours. The purple curves depict the $\bino$ decay length.
To the right of the green dashed curve $\bino$ decays after the freeze-out of $\tbino$. 
} \label{fig:M1mu}
\end{figure}

Figure~\ref{fig:M1mu} shows the SI scattering cross-section in the $(M_1,\,\mu)$-plane for $\tan\beta=5$, $v'/v=3$ and $m_{\tau'}/m_{DM}~=~0.6$,
with the right-handed twin stau mass $m_{\tilde{\tau}_R'}\equiv m_{\tilde{\tau}'}$ determined to obtain $\Omega h^2=0.12$.
The required value of $m_{\tilde{\tau}_R'}$ is larger than the one in figure~\ref{fig:Omega} because of the annihilation of $\tbino$ into $\tau'$ via the mixing with the twin higgsino.  For $\mu M_1>0$, $|\mu|$ below about 600 GeV is excluded by Xenon1T~\cite{Aprile:2018dbl}. The unconstrained natural parameter space will be probed by future experiments. For example, LZ~\cite{Akerib:2019fml} will probe $\mu$ up to about 3~TeV.  We will comment on the blind spot in DD~\cite{Cheung:2012qy} for $\mu M_1< 0$ later.

\begin{figure}[t]
\begin{center}
    \includegraphics[width=0.48\textwidth]{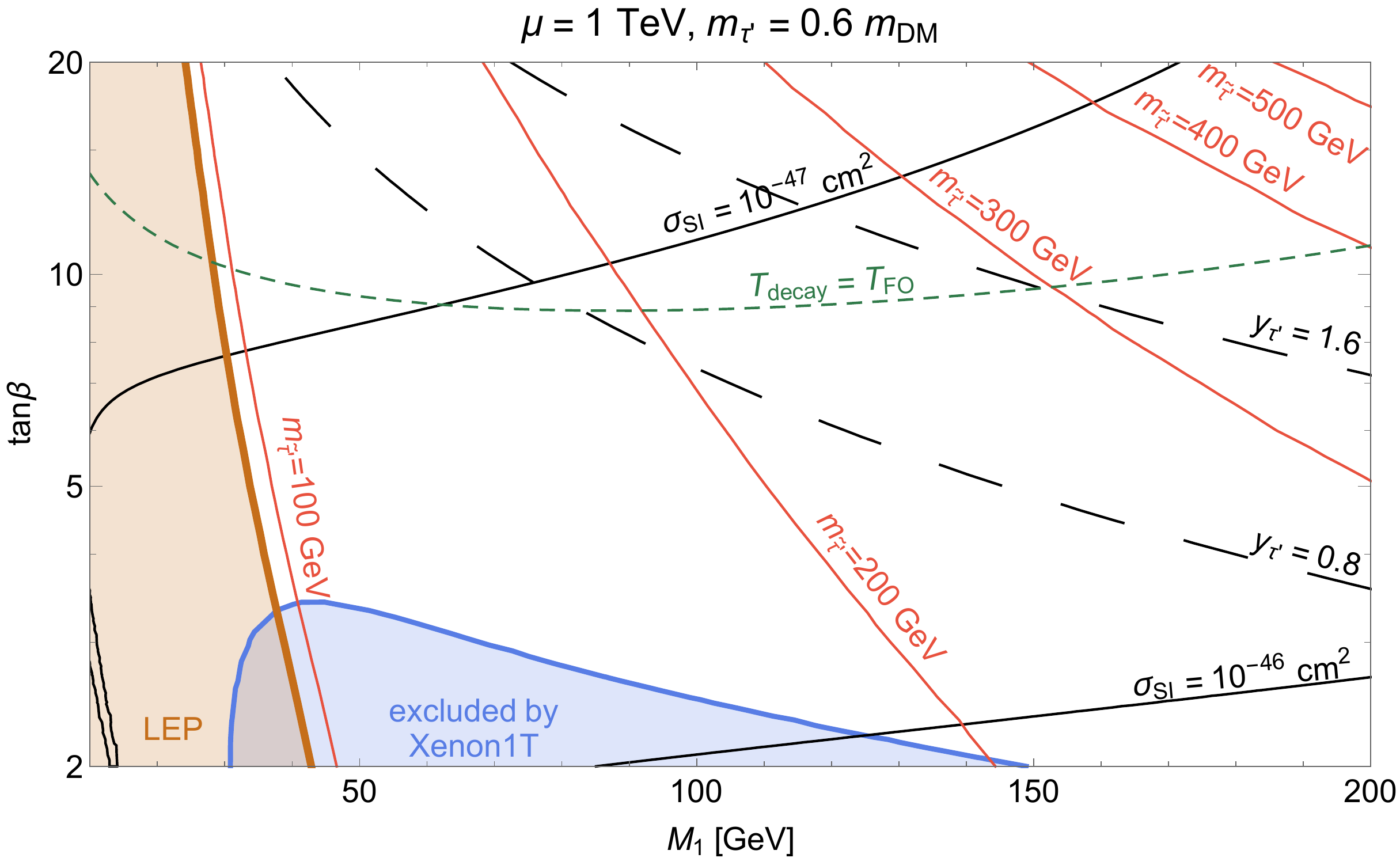}
  \end{center} 
\caption{The same as in figure~\ref{fig:M1mu} but in the $(M_1,\,\tan\beta)$-plane for $\mu=1$~TeV, $v'/v=3$ and $m_{\tau'}/m_{\rm DM}=0.6$. Above the green dashed curve, $\bino$ decays after the freeze-out of $\tbino$. } \label{fig:M1tanb}
\end{figure}

Eq.~\eqref{eq:chLSP} shows that generically the SI scattering cross-section is smaller for larger $\tan\beta$. However, $\tan\beta$ is bounded from above by the perturbativity of the twin tau Yukawa coupling $y_{\tau'}$.
A crucial feature of our scenario is that the twin tau mass, which is given by $m_{\tau'}=y_{\tau'}v' \cos\beta$, is relatively large.
Since $v'/v\lesssim4$ is required to keep fine-tuning at the level of 10\%, large $y_{\tau'}$ is necessary to compensate the suppression by large $\tan\beta$. Avoiding a Landau pole below $10^4$ ($10^{16}$)~GeV requires $y_{\tau'}$ at the electroweak scale below about 1.6 (0.8). Figure~\ref{fig:M1tanb} shows that the allowed range of $\tan\beta$ is limited. For a twin tau mass of 90~GeV (which corresponds to $m_{\rm DM}=150$~GeV in figure~\ref{fig:M1tanb}), $\tan\beta$ must be below 10 (5) to keep perturbativity up to $10^4$ ($10^{16}$)~GeV. 

For a large twin tau Yukawa coupling, there are potentially large corrections to the right-handed stau mass.
The correction to the mass by
the mixing between the right-handed and left-handed twin staus is
\begin{align}
\label{eq:corLR}
\left. \Delta m_{\tilde{\tau}_R'}^2 \right|_{\rm tri-linear} \simeq - y_{\tau'}^2 s_\beta^2 \frac{\mu^2 v'^2 }{m_{\tilde{\tau}_L'}^2},
\end{align}
where $m_{\tilde{\tau}_L'}$ is the twin left-handed stau mass. 
For similar soft masses of the left-handed and right-handed staus this correction may lead to fine-tuning to obtain light right-handed stau.  
The correction is small for large $m_{\tilde{\tau}_L'}$, but then the quantum correction
\begin{align}
\label{eq:corloop}
\left. \Delta m_{\tilde{\tau}_R'}^2 \right|_{\rm quantum} \simeq - \frac{y_{\tau'}^2}{4\pi^2} m_{\tilde{\tau}_L'}^2 
\end{align}
may become large but we find that for a large range of $m_{\tilde{\tau}_L'}$ the fine-tuning in the stau mass is small for natural values of $\mu$ unless $\tan\beta\gtrsim20$  which is already disfavored by the perturbativity of $y_{\tau'}$.

We do not expect any signal in DD via spin-dependent interactions of the twin neutralino with nucleons, since twin neutralinos do not interact with the $Z$ boson which mediates this interaction. This feature is independent of UV completions since the mixing of the $Z$ boson with the twin $Z$ boson must be small to satisfy the electroweak precision tests~\cite{Erler:2009jh}.  Signals in indirect detection are small because $\tbino$ annihilation into SM particles via the neutral Higgs exchange is suppressed. 
Annihilation cross-sections into twin states are sizable but those particles rarely decay to SM states.

{\it Bino-twin bino mass splitting}---%
An important feature affecting the phenomenology in this scenario is the small mass splitting between $\bino$ and $\tbino$. In the limit $|M_1| \ll |\mu|$, from the results for the MSSM~\cite{Badziak:2015qca}, we find
\begin{equation}
\label{eq:massdif}
 \Delta m_{\bino}\equiv m_{\bino} - m_{\tbino} \approx \frac{g_1^2 \left(v'^2-v^2 \right) }{2\mu^2} 
\,\left( \mu\, s_{2\beta}
+M_1 \right),
\end{equation}
where we take $M_1 >0$ without loss of generality, and assume CP symmetry. 
For $M_1 >  - \mu s_{2\beta}$, which we assume throughout this Letter, $\bino$ is heavier than $\tbino$ due to a smaller mixing with the higgsino. The mass splitting tends to zero for larger $|\mu|$ and increases for larger $v'/v$. 

A small $\bino-\tbino$ mass splitting affects the scenario in two ways. First, $\bino$ may be too long-lived and decay after the $\tbino$ freeze-out. The chemical equilibrium between $\bino$ and $\tbino$ is not maintained and $\tbino$ produced by the late decay of  $\bino$   overcloses the universe. Second, even if the $\bino$ lifetime is short enough to avoid the $\tbino$ overabundance, a mass splitting of ${\mathcal O}(5)$\% leads to coannihilation, enhancing the $\tbino$ relic abundance.

The $\bino$ decay width depends on the mixing between the MSSM and twin neutralinos which arises from the UV completion. In SUSY $D$-term TH models~\cite{Badziak:2017syq,Badziak:2017wxn,Badziak:2017kjk}, the mixing between the higgsino and the twin higgsino leads to a coupling between $\bino$, $\tbino$ and the $Z$-boson,
\begin{eqnarray}
g_{BB'Z}  \simeq 
   4\cdot 10^{-5} \left(\frac{g_X}{2}\right)^2\left(\frac{8\,\mathrm{TeV}}{m_{\tilde{X}}}\right)\left(\frac{v'}{3v}\right)^2\left(\frac{\mathrm{TeV}}{\mu}\right)^3\left(\frac{5}{t_\beta}\right)\quad
\end{eqnarray}
where the large $\tan\beta$ and $\mu$ limit is used. Here $g_X$ is the gauge coupling constant of an extra gauge interaction in the $D$-term model, $m_{\tilde X}$ is the mass of the corresponding gaugino.
The decay rate of $\bino$ into $\tbino$ and a pair of SM fermions via an off-shell $Z$-boson is
\begin{eqnarray}
  \Gamma_{\bino}  
 \simeq 2.2\cdot 10^{-16}  \left(\frac{ \Delta m_{\bino}}{7\,\mathrm{GeV}}\right)^5 \left(\frac{g_{BB'Z}}{4\cdot 10^{-5}} \right)^2 \,\mathrm{GeV}.
\end{eqnarray}
In the Supplemental Material more details about the calculation of the bino decay rate are provided. 

In figures~\ref{fig:M1mu} and \ref{fig:M1tanb} we present the region in which $\bino$ decays after the freeze-out of $\tbino$ and overproduces $\tbino$, assuming a $D$-term model with $m_{X}=8$~TeV and $g_X=2$, which are typical values minimizing the fine-tuning of the electroweak scale.
We see in figure~\ref{fig:M1mu} that this does not occur for $\mu \lesssim 1$ TeV, which is anyway required by the naturalness of the electroweak scale.
The $\bino$ decay width can be enhanced for large $v'/v$ but $v'/v\gtrsim4$ leads to fine-tuning worse than 10\%.
Figure~\ref{fig:M1tanb} shows that $\bino$ decays before the freeze-out of $\tbino$ for most of the parameter space where the perturbativity of $y_{\tau'}$ is maintained.
For small mass splitting the scattering $\bino f \leftrightarrow \tbino f $ by $Z$ boson exchange dominates over the decay to maintain the chemical equilibrium. However, this does not expand the parameter space to the region which is already disfavored by the perturbativity of $y_{\tau'}$ or the naturalness of the electroweak scale.
The effects of coannihilation are included in figures~\ref{fig:M1mu} and \ref{fig:M1tanb} but impacts the relic abundance only marginally in the region in which $\bino$ decays before the freeze-out of $\tbino$.
The kinetic and/or soft mass mixing between $\bino$ and $\tbino$ can lead to more efficient chemical equilibration, but we do not investigate this possibility further since the equilibration condition is anyway not constraining.

Let us also comment on the case with $\mu M_1 <0$. Eq.~\eqref{eq:chLSP} shows that the Higgs-LSP coupling vanishes at tree-level for $M_1\approx-\mu s_{2\beta}$  and the SI scattering cross-section is small, which is the so-called blind spot~\cite{Cheung:2012qy} in DD. However, in the blind-spot region the twin stau mass required to achieve $\Omega h^2\approx0.12$ is close or even below 100~GeV so is either excluded by LEP or within the reach of the high-luminosity LHC~\cite{CMS:2018imu}. This is the case for two reasons. First, for $\mu<0$ there is a cancellation in the s-wave annihilation amplitude. Second, in the blind-spot region the $\bino-\tbino$ mass splitting is small, cf.~eq.~\eqref{eq:massdif}, 
so that coannihilation between $\bino$ and $\tbino$ is effective and further suppresses the effective annihilation cross section.    
The strong upper bound on the stau mass may be avoided for a heavier LSP for which annihilations into twin gauge bosons are kinematically allowed but this requires dedicated study which we leave for future work.

{\it LHC phenomenology}---%
The direct-detection, naturalness and perturbativity constraints together with the requirement of the correct relic abundance set an upper bound on the twin stau mass of few hundreds GeV.
The direct production cross-section for a pair of 200~GeV right-handed staus is $\mathcal{O}(10)$~fb~\cite{Fiaschi:2018xdm,SUSYWG} and the LHC searches for ${\tilde \tau}_R \to \tau \tilde\chi_1^0$ are still statistically limited to set meaningful constraints.
The MSSM right-handed stau is generically expected to be somewhat heavier than the twin stau due to the negative loop correction, eq.\eqref{eq:corloop},  and the correction from left-right mixing, eq.\eqref{eq:corLR}, to the latter. 
Nevertheless,  the MSSM right-handed stau is not much heavier than its twin counterpart unless fine-tuning in the twin stau mass is present. 
Thus, the right-handed staus in the preferred mass range may be in the discovery reach of the high-luminosity LHC~\cite{CMS:2018imu}. This is an important complementary probe of this scenario, especially for the $\mu M_1 <0$ case which will not be covered by future DD experiments in the vicinity of the blind spot.

The constraint on the higgsino mass
depends on the decay pattern of charginos and neutralinos. In the minimal scenario that we consider, the higgsino-like chargino $\tilde{H}^{\pm}$ typically decays to $W^{\pm} \tilde{B}$, while the higgsino-like neutralinos $\tilde{H}^0$ decay to $\tilde B$ accompanied by a $Z$ or Higgs boson. For this decay topology the strongest constraint on $\mu$ is set by the ATLAS search for
$\tilde{H}^\pm \tilde{H}^0$
direct production~\cite{Aad:2019vvf} which excludes the higgsino-like chargino mass up to about 500~GeV.
This is weaker than the bound on $\mu$ from Xenon1T but it may get stronger as the LHC collects more data. 

Collider signatures are not exactly the same as in the MSSM because $\bino$ is not stable and eventually decays to $\tbino$ via an off-shell $Z$ boson. Interestingly $\bino$ is typically long-lived with a decay length varying from $\mathcal{O}$(mm) to several meters, as shown in figure~\ref{fig:M1mu}, leading to displaced vertices.
It may be challenging to reconstruct such displaced decays because the typical $\bino-\tbino$ mass splitting is below 10~GeV. However, sensitivity to direct stau and higgsino production may improve by combining cuts from displaced searches with those from usual prompt searches with large missing transverse energy, see~\cite{Allanach:2016pam}.

{\it Discussion}---%
We have investigated the phenomenology of the twin LSP. We focused on the twin bino-like LSP mixing with the twin higgsino and found that it is a promising thermal DM candidate. The twin LSP interacts with SM particles via the Higgs portal, leading to DD signals detectable by future experiments. 
This is the first example of a twin SUSY state playing the role of DM.
It will be interesting to examine other SUSY DM candidates in the twin sector.

\section*{Acknowledgments}
MB would like to thank Satoshi Shirai for useful discussions. MB and GGdC acknowledge the GGI Institute for Theoretical Physics for its hospitality and partial support where some of this work was done. MB and GGdC have been partially supported by the National Science Centre, Poland, under research grant no. 2017/26/D/ST2/00225. KH has been  supported by the Director, Office of Science, Office of High Energy and Nuclear Physics, of the U.S.~Department of Energy under Contract DE-SC0009988 and the Raymond and Beverly Sackler Foundation Fund.

\bibliography{draft}

\end{document}


\title{Natural Twin Neutralino Dark Matter: supplemental material}

\author{Marcin Badziak, Giovanni Grilli di Cortona}
\affiliation{Institute of Theoretical Physics, Faculty of Physics, University of Warsaw, ul.~Pasteura 5, PL--02--093 Warsaw, Poland}
\author{Keisuke Harigaya}
\affiliation{School of Natural Sciences, Institute for Advanced Study, Princeton, New Jersey 08540, USA}

\begin{abstract}

\end{abstract}

\maketitle

In this supplemental material more details of the calculation of the bino decay width in SUSY $D$-term TH models are provided.

In SUSY $D$-term TH models~\cite{Badziak:2017syq,Badziak:2017wxn,Badziak:2017kjk}, the higgsino mixes with the twin higgsino through the gaugino of the gauge multiplet whose D-term generates the $SU(4)$ invariant  quartic coupling. The mixing results in a coupling between $\bino$, $\tbino$ and the $Z$-boson,

\begin{eqnarray}
&&g_{BB'Z}  =\frac{g_2\,g_X^2}{4\, \cos \theta_W\,m_{\tilde{X}}}\sum_{i=1}^4\frac{v\, v'\,s_\beta^2}{m_{\tilde{\chi}_i ^0}-m_{\tilde{\chi}_1^{'0}}}\times\\
&&(N^{-1}_{4i} N^{-1}_{41}-N^{-1}_{3i}N^{-1}_{31})\left(N^{-1}_{4i}-\frac{N^{-1}_{3i}}{t_\beta}\right)\left(N^{'-1}_{41}-\frac{N^{'-1}_{31}}{t_\beta} \right) \nonumber \\
    &&\simeq4\cdot 10^{-5} \left(\frac{g_X}{2}\right)^2\left(\frac{8\,\mathrm{TeV}}{m_{\tilde{X}}}\right)\left(\frac{v'/v}{3}\right)^2\left(\frac{1 \,\mathrm{TeV}}{\mu}\right)^3\left(\frac{5}{t_\beta}\right) \,.\nonumber
\end{eqnarray}
In the approximate equality the large $t_\beta\equiv\tan\beta$ and $\mu$ limit is used. Here $\theta_W$ is the weak mixing angle, $s_\beta\equiv\sin\beta$, $g_2$ is the $SU(2)$ SM gauge coupling,
$g_X$ is the gauge coupling constant of an extra gauge interaction in the $D$-term model, $m_{\tilde X}$ is the mass of the corresponding gaugino, $m_{\tilde{\chi}_i ^0}$ are the neutralino mass eigenvalues, and $N_{ij}^{(')}$  are the mixing matrix elements of the (twin) neutralinos $\tilde{\chi}_i^{(')0}$ defined by
\begin{eqnarray}
\tilde{\chi}_i ^0= N_{ij}\Psi_j \qquad \Psi\equiv (\tilde{B}_{\rm EW},\tilde{W}^0_{\rm EW}, \tilde{H}_{d,{\rm EW}}^0, \tilde{H}_{u,{\rm EW}}^0)^T,
\end{eqnarray}
with the subscripts EW denoting the electroweak eigenstates and $\tilde{\chi}_1 ^0 = \tilde{B}$.
The decay rate of $\bino$ into $\tbino$ and a pair of SM fermions via an off-shell $Z$-boson is
\begin{eqnarray}
  \Gamma_{\bino}  &=& \frac{g_2^2\,  g_{BB'Z}^2}{24 \pi^3 \cos^2\theta_W} \frac{\Delta m_{\bino}^5}{m_Z^4}  \sum_{2m_f < \Delta m_{\bino}}N_f (I_3^f - \sin^2\theta_W q_f)^2  \nonumber\\
 &\simeq& 2.2\cdot 10^{-16}  \left(\frac{ \Delta m_{\bino}}{7\,\mathrm{GeV}}\right)^5 \left(\frac{g_{BB'Z}}{4\cdot 10^{-5}} \right)^2 \,\mathrm{GeV},
\end{eqnarray}
where $N_f$ is $1$ for leptons and $3$ for quarks, $q_f$ the electric charge and $I_3^f$ the isospin.